# Instant Automated Inference of Perceived Mental Stress through Smartphone PPG and Thermal Imaging

Youngjun Cho[1,2], Simon J. Julier[1], Nadia Bianchi-Berthouze[1,2]

[1]Department of Computer Science, University College London, London, WC1E 6BT, UK
[2]Interaction Centre, Faculty of Brain Sciences, University College London, London, WC1E 6BT, UK

Corresponding Author:
Youngjun Cho
UCL Interaction Centre (UCLIC)
Faculty of Brain Sciences
University College London
66 - 72 Gower Street
London
United Kingdom
Phone: +44 (0)20 3108 7177 (x57177)
Email: youngjun.cho@ucl.ac.uk

## Abstract

**Background:** A smartphone is a promising tool for daily cardiovascular measurement and mental stress monitoring. A smartphone camera-based PhotoPlethysmoGraphy (PPG) and a low-cost thermal camera can be used to create cheap, convenient and mobile monitoring systems. However, to ensure reliable monitoring results, a person has to remain still for several minutes while a measurement is being taken. This is very cumbersome and makes its use in real-life mobile situations quite impractical.
**Objective:** We propose a system which combines PPG and thermography with the aim of improving cardiovascular signal quality and capturing stress responses quickly.
**Methods:** Using a smartphone camera with a low cost thermal camera added on, we built a novel system which continuously and reliably measures two different types of cardiovascular events: i) blood volume pulse and ii) vasoconstriction/dilation-induced temperature changes of the nose tip. 17 healthy participants, involved in a series of stress-inducing mental workload tasks, measured their physiological responses to stressors over a short window of time (20 seconds) immediately after each task. Participants reported their level of perceived mental stress using a 10-cm Visual Analogue Scale (VAS). We used normalized K-means clustering to reduce interpersonal differences in the self-reported ratings. For the instant stress inference task, we built novel low-level feature sets representing variability of cardiovascular patterns. We then used the automatic feature learning capability of artificial Neural Networks (NN) to improve the mapping between the extracted set of features and the self-reported ratings. We compared our proposed method with existing hand-engineered features-based machine learning methods.
**Results:** First, we found that the measured PPG signals presented high quality cardiac cyclic information (relative power Signal Quality Index, pSQI: M=0.755, SD=0.068). We also found that the measured thermal changes of the nose tip presented high quality breathing cyclic information and filtering helped extract vasoconstriction/dilation-induced patterns with fewer respiratory effects (respiratory pSQI: from M=0.714 to M=0.157). Second, we found low correlations between the self-reported stress scores and the existing metrics of the two cardiovascular signals (i.e. heart rate variability and thermal directionality metrics) from short measurements, suggesting they were not very dependent upon one another. Third, we tested the performance of the instant perceived stress inference method. The proposed method achieved significantly higher accuracies than existing pre-crafted features based-methods. In addition, the 17-fold Leave-One-Subject-Out (LOSO) cross-validation results showed that combination of both modalities produced higher accuracy in comparison with the use of PPG or thermal imaging only (PPG+Thermal: 78.33%; PPG: 68.53%; Thermal: 58.82%). The multimodal results are comparable to the state-of-the-art automatic stress recognition methods that require long term measurements (usually, at least a period of 2 minutes is required for an accuracy of around 80% from LOSO). Lastly, we explored effects of different widely-used data labeling strategies on the sensitivity of our inference methods. Our results showed the need for separation of and normalization between individual data.
**Conclusions:** Results demonstrate the feasibility of using smartphone-based imaging for instant mental stress recognition. Given that this approach does not need long-term measurements requiring attention and reduced mobility, we believe it is more suitable for mobile mental healthcare solutions in the wild.

## Introduction

Human physiological events are controlled by the actions of the Sympathetic and the ParaSympathetic Nervous Systems (SNS and PSNS). Of the many different types, cardiovascular and respiratory events have been shown to be important for monitoring a person's mental health and stress [1–5]. Recent studies have demonstrated that it is possible to use smartphone RGB cameras to measure Blood Volume Pulse (BVP) [6–10] and mobile thermal cameras attached to a smartphone (or integrated into it, for example, Cat S60) to measure respiratory cycles [11]. These encouraging results suggest that smartphones could become a powerful apparatus for monitoring and supporting mental stress management on a daily basis through biofeedback [12]. Indeed, the combination of RGB and thermal cameras into one device has the potential to provide a very large set of physiological measurements for stress monitoring in our daily life. Smartphone apps with such capabilities are increasingly desired as possible tools for facilitating stress self-management [13–15] as people are often unaware of their level of stress and of being stress-sensitive to particular situations (e.g. chronic pain can cause a fear of movement [16]). There is also strong interest within the industry in complementing typically used questionnaires in order to enable improved assessment of wellbeing with personnel as well as revisiting work plans and work environments [17]. Given their size and mobility, such sensors could be embedded into employees' aids for ease of use. While these low-cost sensors are still not perfect, the literature shows that their reliability is increasing, and we are contributing to this body of work. At the same time, we hope that our work contributes to the literature in general using these signals as stress measures [18–20]. In this paper, we aim to focus on two important cardiovascular events that can be captured by low cost, low resolution sensors: cardiac cyclic events with smartphone PPG, and vasoconstriction/dilation-induced nose tip temperature dynamics with a low cost thermal camera. In particular, we investigate how to instantly capture stress-induced variability of such physiological patterns.

Heart Rate Variability (HRV) is the time series of variation in heartbeats. It has been used to measure a person's mental stress [4,18,20–25]. HRV's popularity arises from the fact that it has been shown to abstract information about the sympathovagal balance between the SNS and PSNS. When confronted with a stressor, the autonomic nervous system can produce a sequence of fight-or-flight responses [1]. These manifest themselves as alternations of accelerated and decelerated cardiovascular patterns [1,26]. To characterize the HRV, various authors [4,21,22,27] have proposed a variety of hand-crafted HRV metrics that are computed over the time intervals between heartbeats. Although most of the HRV metrics were originally built based on the R-R intervals from ECG (Electrocardiogram) measurements [28], the metrics have been applied to the P-P intervals from PPG measuring blood volume pulse [18,20,25,29]. In the case of PPG, the term Pulse Rate Variability (PRV) or PPG HRV are often used to clarify the different type (even if related) of event measured [26,29–31] with respect to ECG. Amongst the most commonly used are statistical metrics (such as the standard deviation of R-R or P-P intervals) and frequency-band metrics (e.g. the normalized power in a frequency band of interest). In particular, various studies have found that the Low Frequency (LF; 0.04Hz – 0.15Hz) and High Frequency (HF; 0.15Hz – 0.4Hz) bands of the time intervals in heart rates appear to reflect the SNS and PSNS activities [21]. Based on this observation, many studies have proposed to use the LF/HF ratio as a stress indicator [4,22,24,32]. However, the use of such metrics has remained controversial in that they tend to oversimplify physiological phenomenon [33–35]. In particular, a single physiological metric itself does not strongly contribute to automatically detecting a person's stress levels (i.e. machine learning tasks) [33,36]. Hence, multiple HRV metrics-derived features have been used together with those from other physiological activities such as perspiration and respiratory activities for automatically inferring mental stress (e.g. during driving tasks [37] and desk activities [25]). To ensure reliable measurements with such features, a relatively long term window of data (several minutes to a few hours) must also be used [25,36]. Although this is acceptable in specialist settings or with medical devices, it is highly inconvenient in the real world with unstructured settings using low cost devices (in particular, the PPG). For example, if smartphone-based finger PPG was to be used, a user would have to continuously make sure their finger is held stably in front of the camera. Another issue is that changes in ambient light levels, as a user moves around, can corrupt long-term measurements.

Another documented cardiovascular event which happens as a reaction to mental stressors is vasoconstriction of blood vessels in a person's nasal peripheral tissues [38,39]. This causes blood flow to drop, resulting in a decrease in temperature which can be detected by monitoring the temperature of the nose tip. The study in [40] found that a contact-based multi-channel thermistor was able to detect a significant decrease in temperature of the nasal area as relative to the forehead in mentally stressful conditions. The same result has been repeatedly reported from the use of thermal imaging in mental stress induction studies [38,41], indicating that the thermal directionality (i.e. temperature drop) can be a potential barometer of mental stress. However, studies show similar limitations as they require keeping the head still (often authors use a chinrest). In addition, they also require to measure baseline temperatures to compute the thermal direction which may limit its use in real-life applications [42,43]. In this work, we address the former issue by using a state-of-the-art tracking method [11]. Furthermore, we rely only on the instant measurement with the area of interest (nose tip) to address the latter.

The reason for proposing the use of two sensors in this study rather than just one is that, despite the potential of thermal imaging in measuring BVP [44], its accuracy is low and its ability in measuring P-P intervals has not been yet validated. Instead, camera-based PPG has been shown to be more reliable [9,45] and can be used simultaneously with thermal imaging, possibly compensating each unimodal performance in inference tasks. In addition, the use of finger PPG and thermal camera raises much less privacy concerns than RGB-based facial analysis (i.e. remote PPG [8]). Furthermore, the use of multiple measurements increases reliability of stress monitoring. Finally, even if not investigated in this paper, low cost thermal imaging could provide further measurements of stress-related phenomena (e.g. respiration rate [11,36] has already shown to be possible with a mobile thermal camera, and possibly sweat [46]) to provide a wide battery of cues for reliable assessment.

Rather than focusing on all possible physiological signals that could be later added, this paper investigates the possibility to build a fast stress recognition system that only requires a very short time window of PPG and thermal measurements. This is to ensure the possible use in real-life ubiquitous situations. In particular, we contribute to the literature on four fronts. First, we propose new preprocessing techniques to enhance the quality of the signals that are extracted from both the smartphone-based PPG and thermal camera, and to reliably produce P-P intervals and thermal variability data as low-level features. This is particularly important when working with ultra-short measurements [47]. Second, we explore correlations between currently used metrics from thermal and PPG signals over a short period of time and self-reported stress scores. Third, instead of using the existing metrics as high-level features, we propose to use the low-level features and let artificial neural networks (NNs) to learn informative high-level ones themselves. We evaluate the approach on a multimodal dataset purposely collected for this study. Finally, we further investigate sensitivities of different labeling strategies from self-reported stress scores within the perceived stress recognition performance.

## Methods

This section presents a method that enables quick inference of a person's perceived stress level using smartphone-integrated PPG and thermography. We call these measurements *instant measurements* to differentiate them from the *short measurements* (typically between 2min and 5min) which have been previously defined in the literature [47].
First, we describe software we implemented. This includes a recording set-up and a set of techniques to produce reliable PPG-derived HRV profiles and sequential nose tip thermal variations (called hereafter the *thermal variability sequence*) from the thermal imaging sensor. We then introduce our study protocol to induce different levels of mental stress and collect short sequences (20s) of cardiac pulse-related and thermal events together with self-reports of perceived mental stress scores. Third, we extract low-level (one-dimensional P-P intervals and thermal variability sequences) and high-level hand-engineered features, comparing the performance of our system over the two sets of features and sensor modalities. We conclude by comparing our approach to data labeling with standard approaches to discuss the effect of inter-subjective variability in reporting stress scores.

### Toward Smartphone as a Reliable Multiple Cardiovascular Measure

The main cardiovascular sensing channels of this work are the rear RGB camera of a mobile phone (LG Nexus 5) and a low-cost thermal camera (FLIR One 2G) attached to the phone. Figure 1 shows the smartphone with the attached thermal camera, the required finger placement and light emission for PPG, and the physiological measurement interface.

Although the smartphone-imaging-based PPG measurement can be performed in either a contact [6,7] or a contactless manner [8], in our work we only focus on a contact-based imaging PPG. The reason is based upon previously repeated investigations within clinical studies [6,10] reporting its high accuracy. In addition, given that a normal RGB camera is only sensitive to a narrow electromagnetic spectral range of visible light in the so-called visible spectrum [48], adequate lighting is required before it can be used as a PPG sensor. Hence, a light emission from the rear flash LED is used and a user is required to hold the smartphone body and place his/her finger over both the back camera and flash light (Figure 1a,b). Unfortunately, the use of the back flash limits the duration of the measurements in some devices since its heat can potentially burn a person's skin. As shown in Figure 2, a large amount of heat is produced by the LED emission from the chosen smartphone (LG Nexus 5) in just 25-30 seconds of operation. A similar amount of heat was observed from another mobile phone (Samsung Galaxy 6 in Figure 1b). Since temperatures above 50°C are potentially damaging to human skin tissues (e.g. skin erythema could occur from 25 seconds heating at 51.07°C [49]), we limit the cardiovascular measurement to a 20 second time period. This is also the required minimum duration for obtaining valid HRV metrics values (in particular, LF/HF [47]).

To capture a time series of apparent thermal sequences, we developed bespoke recording software using the FLIR One library. The interface is shown in Figure 1c. Considering the thermal properties of human skin, the emissivity of the thermal imaging sensor was fixed at 0.98 [50]. As the thermal imaging system does not guarantee a consistent frame rate [48], the recording interface stores the time stamp with each image frame.

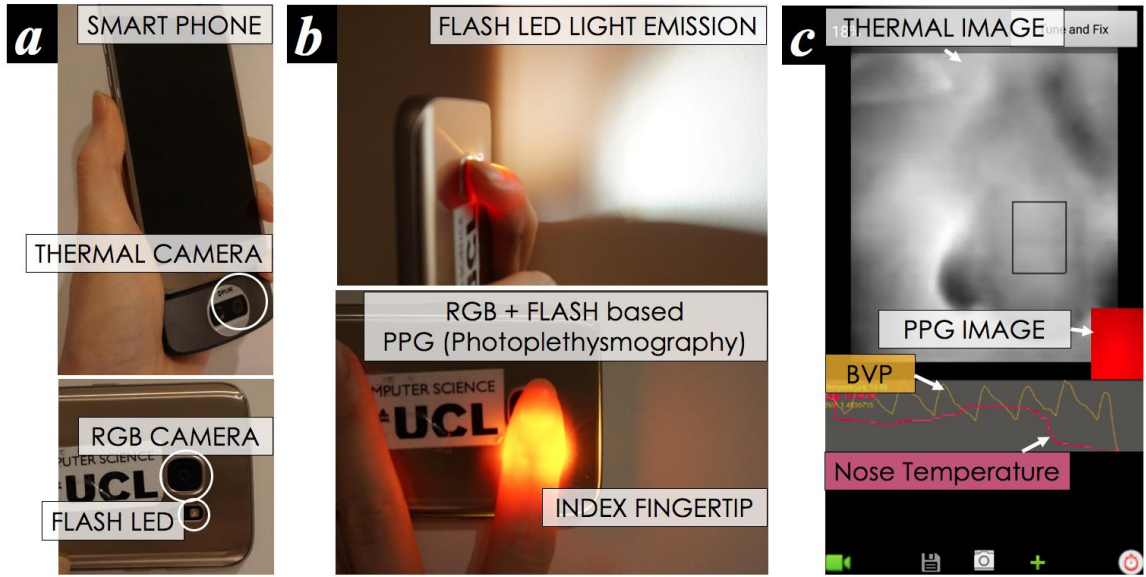

Figure 1. Smartphone RGB and thermal camera based physiological measurement: (a) a smartphone with an add-on thermal camera, (b) flash LED emission and finger placement for PPG measurement, (c) designed software interface to collect BVP and 1D thermal signature from the nose.

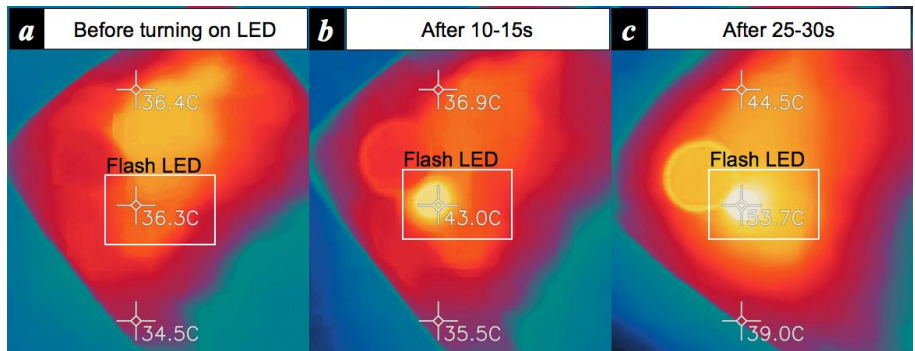

Figure 2. Heat produced by the rear flash LED of a smartphone (LG Nexus 5) measured by a thermal camera (FLIR One): (a) before turning on the LED (36.3°C), (b) after 10-15 seconds (43°C), and (c) after 25-30 seconds (53.7°C).

*BVP and P-P Interval Estimation through PPG Imaging*
Figure 3 summarizes the approach we use to extract BVP and P-P intervals through the smartphone imaging PPG. Following [6,7,10], our method estimates the BVP signals by capturing subtle color variations associated with light absorptivity patterns of hemoglobin in the capillaries of a person's skin. However, rather than using average values of the pixels of the red (or green) channel to estimate the BVP value (which is the most widely used method [6,7,9]), we propose to use the temporal variations in spatial Shannon's entropy [51] of sequential R-channel images as raw BVP signals. This is due to averaging which tends to ignore fairly small but important variations in color distribution [11]. The estimated BVP value $B_t(X)$ at a given time *t* can be expressed as:

$$B_t(X) \equiv -\sum_{(i,j)} p(x_{i,j}) \log_2 p(x_{i,j}) \qquad (1)$$

where $x_{i,j}$ is the brightness of pixel(*i,j*) and $p(x_{i,j})$ is the probability distribution which is generally estimated using a grayscale histogram in image analysis [52] (here, for the R channel).

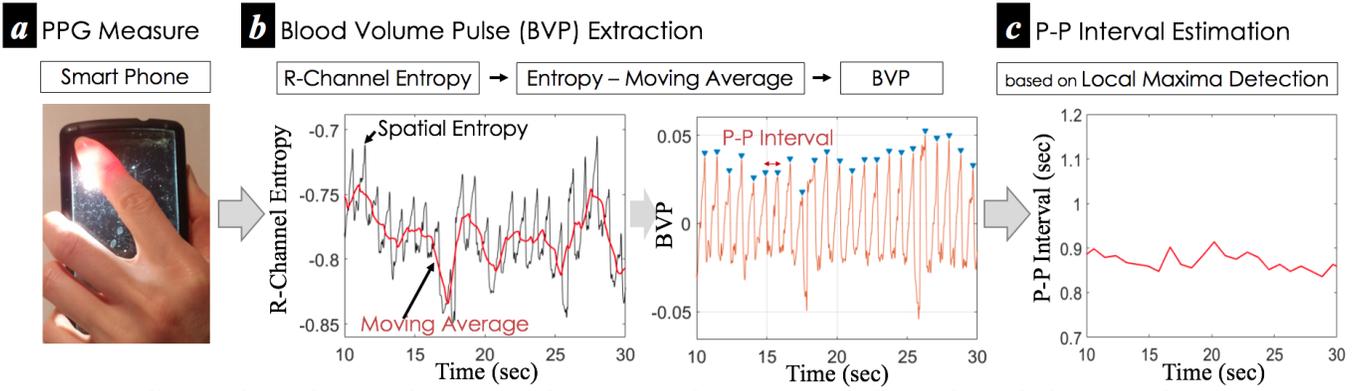

Figure 3. Overall procedure of BVP and P-P interval estimation from a person's finger through the smartphone-imaging PPG. See text for details.

As our interest is in measuring raw P-P intervals from PPG signals, we used a simple signal processing technique to create similar amplitudes of each peak of BVP which helps detect peaks for measuring the time interval (i.e. P-P interval) between the peaks. This was done by the subtraction of the *k*-sample moving average signals from the raw entropy signal (Figure 3b) which can be expressed by

$$\hat{B}_t = B_t - \frac{1}{k}\sum_{i=t-k}^{t-1} B_i .\tag{2}$$

Because a high sampling rate produces a higher sensitivity of the P-P intervals [53], we up-sampled the raw sequences to 256 Hz with spline interpolation and used a 1s moving average to smooth heartbeat induced variations within the duration where at least one heartbeat of a normal person is expected to appear [54]. Finally, we used the simple local maxima detection [55] with a 0.5 second sliding window to recover P-P intervals (Figure 3c).

*Continuous Extraction of Nose Tip Thermal Variability Sequence*
To extract the 1D sequential nose tip thermal changes, our approach uses the three computational steps shown in Figure 4. These are: i) nose-tip Region of Interest (ROI) tracking, ii) breathing artifact reduction, and iii) post processing for extracting low-level features representing thermal variability.

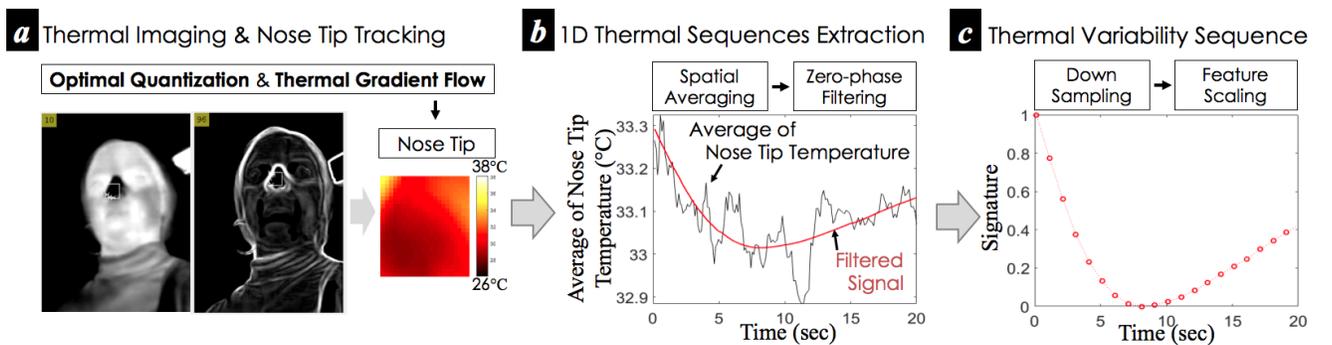

Figure 4. Overall procedure of the extraction of one-dimensional thermal variability signature from a person's nose tip through smartphone thermal imaging.

For ROI tracking, we can take advantage of recent advances in thermal ROI-tracking techniques which help minimize the effects of motion artefacts and thermal environmental changes. In particular, we used the Optimal Quantization and Thermal Gradient Flow methods (Figure 4a) introduced in [11]. Through the use of these techniques, we can continuously extract a spatial average temperature sequence over the ROI. As breathing causes thermal changes in the area close to the nose tip (see Figure 4b), we need to remove such effects from the ROI for reliable measurements. This is necessary despite the fact that breathing dynamics are significant indicators of mental stress [3,36] in itself. For this, we propose to use a low-pass filter with a cutoff frequency lower than the normal range of breathing rates of healthy people (e.g. 0.1-0.85Hz in [11]). As a thermal directional change is a relatively slow physiological event [56], we set this to 0.08Hz which is lower than the low boundary.

For the implementation, we used a zero-phase filtering (seventh-order, Butterworth) to avoid a phase-shifted result. Finally, we computed the thermal variability sequences of the nose tip (Figure 4c) by down-sampling with a linear interpolation and feature scaling the signal. Here, down-sampling (1Hz) is used to address the unsteady frame rate of the thermal camera and to compute successive temperature differences sampled at regular temporal points. Feature scaling (Figure 4c) was applied to minimize the effect of different levels of nasal temperatures across participants and sessions and to explore the thermal temporal variability within short-term data. As this new method helps extract nose tip thermal variability sequences continuously, it can produce richer feature sets in comparison with earlier methods [38,40,41]. In turn, this could possibly provide useful information, even from an instant measurement, contributing to the automatic inference of a person's stress.

## Data Collection Protocol

A data collection study was carried out to gather physiological data from participants during different tasks that induced different levels of mental load. The data collection protocol is described below.

### *Participants*

A total of 17 healthy adults (mean age 29.82 years, SD=12.02; 9 female) of varying ethnicities and different skin tones (pale white to black) were recruited from the University College London and non-research community through the UCL psychology subject pool system. Participants completed prescreening through the system which was designed to exclude participants with any history of psychiatric disorders or medicine intakes which may influence their physiological signatures. Each participant was given the information sheet, asked to provide a signed consent to take part in the study and to fill in the demographics form prior to the start of data acquisition. The study was conducted in a quiet lab room with no distractions. Participants were informed that they could stop the study at any time if they felt uncomfortable. Only one experimenter was present in the room during the data collection but kept his distance from the participant (further than 1.5 m). We compensated each participant with an £8 Amazon voucher after completion of the study. The experimental protocol was approved by the Ethics Committee of the University College London Interaction Centre (ID Number: STAFF/1011/005).

### *Task Structure and Instant Measurements of Lasting Stress-induced Physiological Events*

We designed a stress induction study protocol to collect physiological data and subjective self-reports in association with mental stress levels due to mental load [1,57]. From the literature on mental stress induction studies in psychology, neuroscience and affective computing (e.g. [2,25,58,59]), we chose two cognitive-load induction tasks – the Stroop Color Word test [60] and the Mathematical Serial Subtraction test [61]. These tests were selected as they have been shown in various studies to induce mental stress by increasing cognitive load. They have also been used in other thermal imaging studies [39,41]. Each task was divided into two sub-tasks with varying difficulty levels so as to elicit different stress levels (easy and hard: Se – Stroop easy, Sh – Stroop hard, Me – Math easy, Mh- Math hard) and each sub-task was counterbalanced in a Latin squared design as in [36]. Between sub-tasks, we added a break period encouraging participants to fully recover (without any measurements, constraints) so as to avoid potential effects from previous sessions.

Although it has been shown that the Stroop and Math tasks lead to cognitive overload [2,59], they are limited in the amount of stress they induce due to the lack of psychosocial stressors or other stressors [2,62]. Hence, following [2,40,59,62], we also introduce further stressors: a) *social evaluative threats* (close observation and assessment of a person's performance [2,62]), b) *time pressure* (e.g. 1.5 second limitation for each Stroop question [59]), and c) *loud sound feedback*, in particular, an unpleasant sound for wrong answers [40].

As described above, heat caused by the use of the smartphone PPG limited our data gathering to a 20 second window immediately after each task. The aim is to capture the cardiovascular changes related to stress responses and their dynamics immediately after the stressor has ended instead of measuring the signals during each task (Figure 5). Overall, this study protocol consisted of:

- waiting in the corridor, introduction and entering the study room (5-10 min)
- information/consent/demographics forms filled in (5-10 min)

Session 1
- [Rest 1] **sitting, resting (5 min)**
- *20s measurement and self-reporting of perceived stress (1-2 min)*
- [Task 1] **Stroop Test 1** (5 min)
- *20s measurement and self-reporting of perceived stress (1-2 min)*
- break (5 min)
- [Task 2] **Stroop Test 2** (5 min)
- *20s measurement and self-reporting of perceived stress (1-2 min)*
- break (3 min)

Session 2
- [Rest 2] **sitting, resting** (5 min)
- *20s measurement and self-reporting of perceived stress (1-2 min)*
- [Task 3] **Math Test 1** (5 min)
- *20s measurement and self-reporting of perceived stress (1-2 min)*
- break (5 min)
- [Task 4] **Math Test 2** (5 min)
- *20s measurement and self-reporting of perceived stress (1-2 min)*
- break (5 min)

Closing
- wrap-up and participant's feedback (5-20 min)

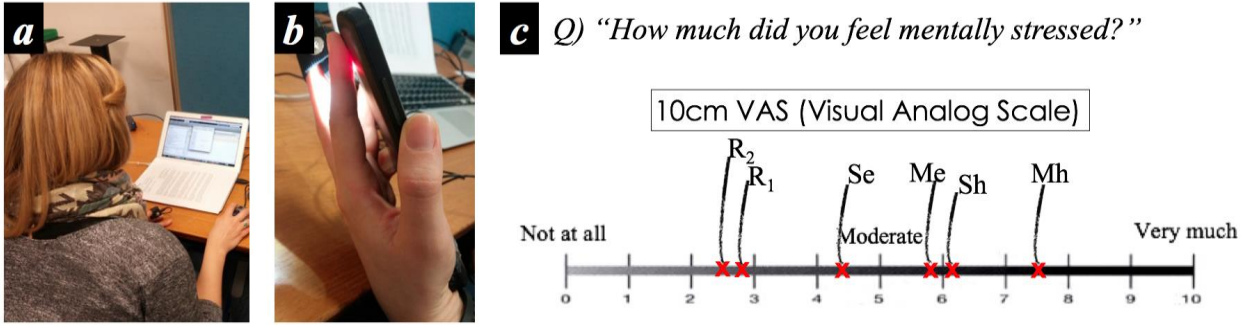

Figure 5. Experimental setup and self-report question: (a) during each stress-induction task session, (b) 20 second physiology measurement after sessions, and (c) 10cm VAS based questionnaire ($R_1$, $R_2$: Rest from Session 1 and 2, Se: Stroop easy, Sh: Stroop hard, Me: Math easy, Mh: Math hard). The red marks (*x*) represent an example of self-reported scores of one participant over the different tasks. The task labels have been added by the researchers for the purpose of this figure.

*Measuring and Self-Report of Perceived Mental Stress*
For the 20s physiological measurements, the person was asked to hold their index finger on the smartphone RGB camera while keeping the smartphone add-on thermal camera facing their nose, as shown in Figure 5b. After each 20 second physiological measurement, all participants were asked to answer a questionnaire about their perceived level of mental stress. We used a 10-cm Visual Analogue Scale (VAS), which allows participants to answer on an analog basis (continuous) to avoid non-parametric properties [63,64]. The question asked was "*How much did you feel mentally stressed?*" (ranging from 0, not at all, to 10, very much). Only one VAS straight line was used for each participant to self-report his/her perceived stress levels across all tasks and sessions. This is to help participants easily compare stress scores they report between sessions as shown in Figure 5c. This approach combines a numerical approach to self-reporting with a ranking one, as ranking is generally more reliable than simple quantization of a subjective state [65–67]. The labels in Figure 5c have been added to the figure by the researcher to clarify their reference to each of the tasks ($R_1$, $R_2$: Rest from Session 1 and 2, Se: Stroop easy, Sh: Stroop hard, Me: Math easy, Mh: Math hard).

## Automatic Inference of Perceived Mental Stress from Instant Measurement

*Low-Level and High-Level Features from Cardiovascular Events*
The 20 second-cardiovascular measurement with the developed interface (Figure 1c, 5b) simultaneously produces the following *signals*:
  a) one-dimensional P-P intervals
  b) one-dimensional thermal variability sequence

We take the P-P intervals (Figure 3c) and thermal variability sequence (Figure 4c) as *low-level* features representing each modality throughout this paper.

In order to evaluate the effectiveness of our approach against standard approaches, we also extracted high-level engineered features for both BVP and nose tip temperature variations as the evaluation benchmark for our approach. We followed earlier studies on stress inference using HRV metrics as the features [25,37,68,69] (in our case, PPG-derived HRV; for readability, hereafter simply called PRV), although we excluded features directly from HR given its minor role repeatedly found in stress inference studies (e.g. [25]). After the pre-processing method described above, we extracted the following PRV features:

a) PRV F1: LF Power
b) PRV F2: HF Power
c) PRV F3: LF/HF ratio
d) PRV F4: SDPP (Standard Deviation of P-P intervals)
e) PRV F5: RMSSD (Root Mean Square of the Successive Differences of P-P intervals)
f) PRV F6: pPP50 (proportion of the number of the successive differences of P-P intervals greater than 50ms *of the total number of the intervals*)

As for *high-level* features representing the nose tip thermal signature, we used the most primarily used feature in the literature [38,40–42]:

g) Nose temperature F1: TD (Temperature Difference between data from the start and the end)

Additionally, we extracted basic statistical features from the processed thermal variability sequence, similarly to SDPP from the P-P intervals:

h) Nose temperature F2: SDSTV (Standard Deviation of the Successive differences of the Thermal Variability sequence)
i) Nose temperature F3: SDTV (Standard Deviation of the Thermal Variability sequence)

The sliding window was not used to extract these features given the short period of time over which they were measured.

*Labeling Strategy and Machine Learning Classifiers*

Given the focus on automated inference of a person's perceived stress level, the labeling of self-reported stress scores is an important step. However, interpersonal variability has been repeatedly found from self-reports of perceived mental stress [24,36,70]. This is a key issue which must be addressed if we are to create automatic stress recognition systems that can generalize across people. Following our earlier work [36], we use the normalized K-means clustering technique to label the measured events, as the K-means has been shown to be effective in handling self-reported data [71]. In detail, all perceived stress scores collected from each participant are normalized through feature scaling that identifies the minimum and maximum scores for a participant and rescales all the scores so the range is the same across all participants. Then, the K-means algorithm (k=3) is used to group the participants' VAS scores into three levels of perceived stress scores corresponding to "None or low stress", "Moderate" and "Very high" on the VAS we used (see Figure 5c). In this paper, we focus on discriminating between two levels of stress, *No-stress* and *Stress* given the limited amount of data for a more refined discrimination. Hence, a third step is required. We split the labels into two groups: the *No-Stress* group referring to the K-mean "None or low stress scores" cluster and the *Stress* group containing both the K-mean "Moderate" and "Very high" score clusters. Two obtained labelled groups are hence used to label the related physiological signatures from each 20s window (L1).

Furthermore, we explored the possible effect of different data labeling strategies: a) L2: combining the first and second K-means clusters (from k=3) into No-stress by contrast with L1, b) L3: K-means with k=2, and lastly, c) L4: the original stress scores divided by directly dividing the VAS scale intro three equal sections and then combining the "moderate" and "Very high" stress classes into one, i.e. "Not at all" and "Moderate + Very high" (threshold at point 3.334 on the VAS scale in Figure 5c). The aim of L2 and L3 was to understand the sensitivity of our approach in separating the moderate level of stress with the other two classes. L4 was used as a way to compare with more standard techniques used in the field [72].

Two machine learning algorithms were tested. First, we used a single hidden-layer Neural Network (NN) which is suitable to work with low-level features (i.e. P-P intervals and thermal variability vectors) capturing their temporal dynamics. The use of artificial NNs can empower automatic learning of informative physiological features with back-propagation to repeatedly tune internal parameters to let the features emerge from the data (this is also called representation learning). Second, with the high-level engineered features, we used the k-Nearest Neighbor classifier (denoted as kNN, k=1) as a benchmark stress inference model given that this is typically used in this area [69]. By choosing this second algorithm, we aim to assess the limitations of the use of handcrafted features which may simplify a person's dynamic physiological events, and in turn possibly miss out some fast, informative moments. In particular, in the case of instant measurements (short period of time), this cannot be compensated by the use of a sliding window producing sequential feature values (e.g. a 120 seconds sliding window used in [25] to continuously produce PRV features during a 180 second task session).

For the implementation of NNs, we tested two sizes of hidden layer nodes: a) small (n=80, NN1) and b) large (n=260, NN2) – each node size was empirically chosen. The mean and standard deviation of the training dataset were used to normalize both the training and testing dataset. The sigmoid was used as an activation function. In the training process, a fixed learning rate of 0.5 was used for 100 epochs.

## Results

In this section we evaluate our proposed approach. First, we report the statistical analysis of the collected data. Second, we discuss the recognition performance of our system over the different modalities and types of features. Finally, we compare the results for the different labeling approaches.

### Reliability of Measured Physiological Patterns

First of all, we tested the reliability of the physiological measurements. From the 17 participants, we collected 102 sets of the estimated BVP signals, P-P intervals and thermal variability sequences from 20s instant measurements taken after each Stroop and Math task and after each resting session. However, 2 sets of data were not recorded due to phone battery issues at the end of one experiment, and 1 set was not recorded as one participant clicked the turn-off button on the phone by mistake. 6 further sets had to be discarded because some participants' nose was not visible on thermal images (nose outside of the range of view due to sudden severe coughing during the 20sec, or because of head turned towards the experimenter, or the nose was covered by a person's hand). Although these disturbances were often transient, they meant that data could not be collected within the 20s immediately following the end of the stressor. An analysis of the thermal data from Rest 1 also showed some extreme patterns in the nose tip temperature (e.g. sudden increase in temperature). This may be explained by the fact that the experiment was conducted during the winter and temperatures outside of the experimental room were often significantly lower. This included both outdoors, and indoors in the corridor where the participants waited for the experiment. Despite the temperature changes, the Rest 1 data was kept in the dataset. A total of 93 sets were used for the study.

As the measurement capability of smartphone PPG has previously been thoroughly investigated in earlier studies (e.g. [6,9,10]), we only tested the reliability of the cardiac pulse signals measured with our approach and compared it with the mean brightness intensity-based method, which has been dominantly used [6,7,9]. For this, we used the relative power Signal Quality Index (pSQI), which is to assess the strength of physiological signals in a frequency range of interest, as a measure of quality [11,53,73,74]. The pSQI for the BVP signals can be expressed by:

$$P(\hat{f}_{min} \leq f \leq \hat{f}_{max}) \cong \frac{\int_{\hat{f}_{min}}^{\hat{f}_{max}} S_{\hat{B}}(f)df}{\int_{total} S_{\hat{B}}(f)df} \quad (3)$$

where $0 \leq P \leq 1$, $S_{\hat{B}}$ is the power spectral density of BVP signals (in our case, $\hat{B}$ in Eq. (2)), and $\hat{f}_{min}$, $\hat{f}_{max}$ are the lower and upper boundary of expected HRs, respectively. Here, we set the expected HR range to [0.8Hz (48bpm), 2.0Hz (120bpm)] given that HRs of healthy adults mostly fall into this range [54]. To minimize effects of the baseline wander and high-frequency noise on this signal quality test [6,74], we used band-pass filtered BVP signals (0.7-4.0Hz) as in [6]. Figure 6 shows the better quality of the estimated BVP signals $\hat{B}$ from the proposed method (Eq. (2)) than that from the mean intensity method (*Proposed*: M=0.755, SD=0.068; *Traditional:* M=0.692, SD=0.075).

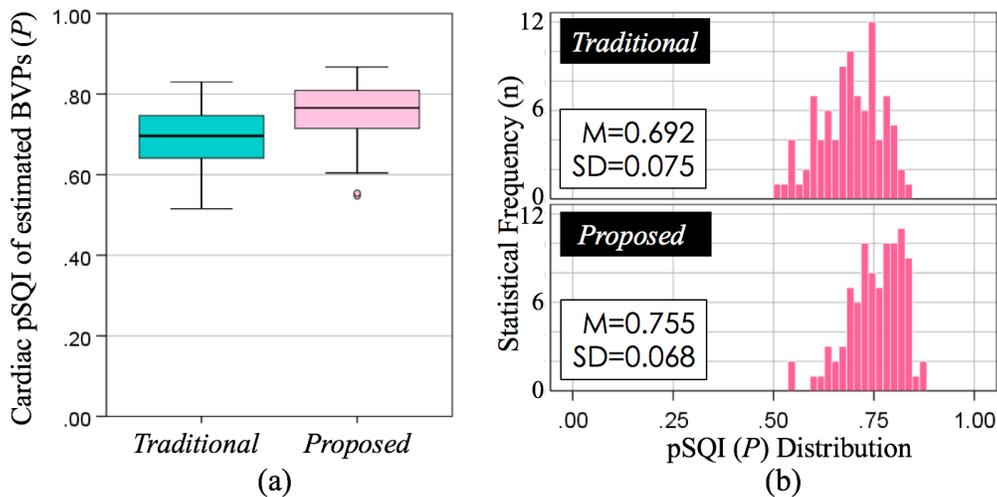

Figure 6. Signal extraction quality comparison of our spatial entropy-based method (Eq.(2)) with the mean intensity approach [6,7,9] by using pSQI: (a) box plot, (b) histogram.

Figure 7a shows examples of thermal images taken from the participants during the data collection study. From our observations, we found that respiration influences the nasal tip temperature measurement in some cases. For instance, in Figure 7b, thermal images of a person's nose tip surface which were sequentially captured show that inhaled air changed the nose tip temperature. Hence, we tested how much participants' respiratory cycled events affected the nose tip temperature measurements by using the pSQI in Eq. (3) with the expected respiratory rate of interest (from 0.1Hz to 0.85Hz) as used in [11]. Figure 7c demonstrates how the measured nose tip temperatures involved respiratory cyclic patterns (respiratory pSQI: M=0.714, SD=0.163), indicating that such affected temperature patterns may lead to wrong stress level classification. On the other hand, the processing technique we propose to use (Figure 4b) instead led to reducing respiratory artifacts on the measurement (respiratory pSQI: M=0.157, Sd=0.091).

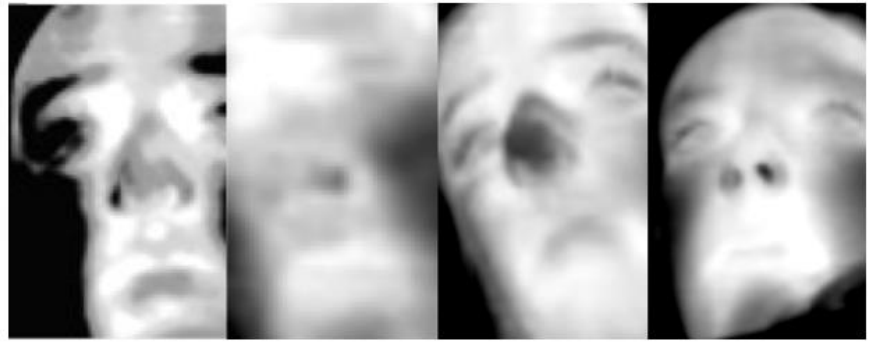

(a) Examples of Thermal Images collected from the 20s measurement

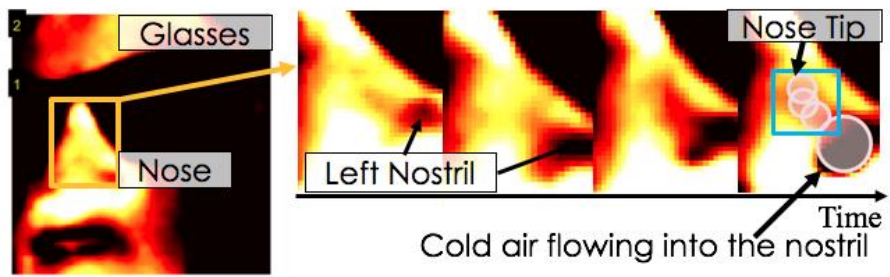

(b) Nose Tip Temperature affected by Breathing Cycles (Inhalation Phase)

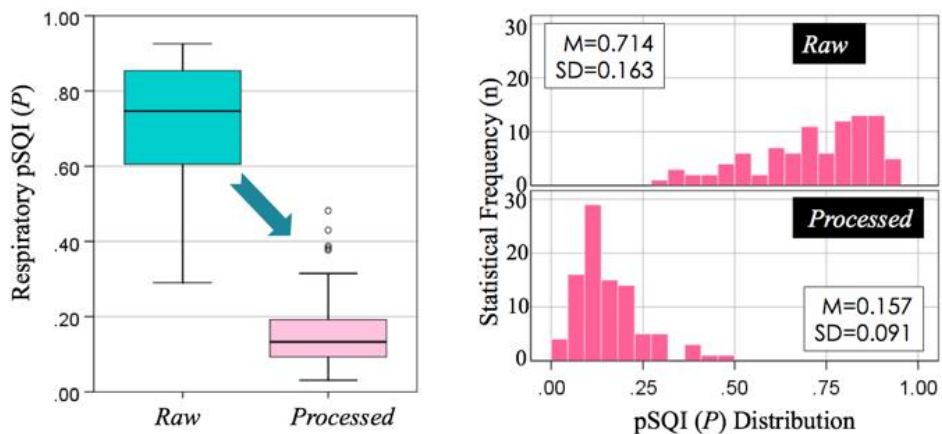

(c) Respiratory pSQI ($P$) comparison

Figure 7. A person's respiratory activity influences the nasal tip temperature: (a) examples of thermal images from participants (view angles were not constrained), (b) the nasal temperature changes during inhalation (yellow: warmer, red: moderate, black: colder), and (c) respiratory signal quality test using pSQI.

## Self-Reported Stress Ratings and Hand-engineered Metrics

An important step was the analysis and possible normalization of the self-reported stress scores. The boxplot in Figure 8a shows the distribution of the self-reported scores over the resting periods and the different sessions and tasks. It is clear that the stress elicitation procedures did, overall, produce the wanted levels of stress with the hard sessions scoring higher than the easy sessions and the latter scoring higher than the resting periods (Rest from Session 1: M=1.49, SD=1.94; Rest from Session 2: M=1.30, SD=1.26; Stroop Easy: M=2.17, SD=1.46; Math Easy: M=2.66, SD=1.80; Stroop Hard: M=3.92, SD=2.11; Math Hard: M=5.17, SD=2.55) despite two outliers. However, the wide boxplots also show inter-subject variability in self-reporting. In addition, the ranges (maximum - minimum) in scores for each participant differ quite highly (Maximum range: 8.75, Minimum range: 1.5, Mean: 4.7, Std: 2.1) further suggesting the need for normalization of the scores.

Therefore, we normalized the data for each participant with respect to their range of scores over all the sessions. Figure 8b shows the original data and Figure 8c shows the normalized data. The normalization helps to identify two main modes in the score distributions suggesting the presence of two main clusters of stress levels. Given the subjectivity of stress ratings and the limited amount of data sets to carry a multi-level model, in this paper we focused on binary classification of perceived mental stress: no/low stress vs. medium/high (or very high) stress. The K-means separation between the two clusters is represented by each different color in Figure 8c.

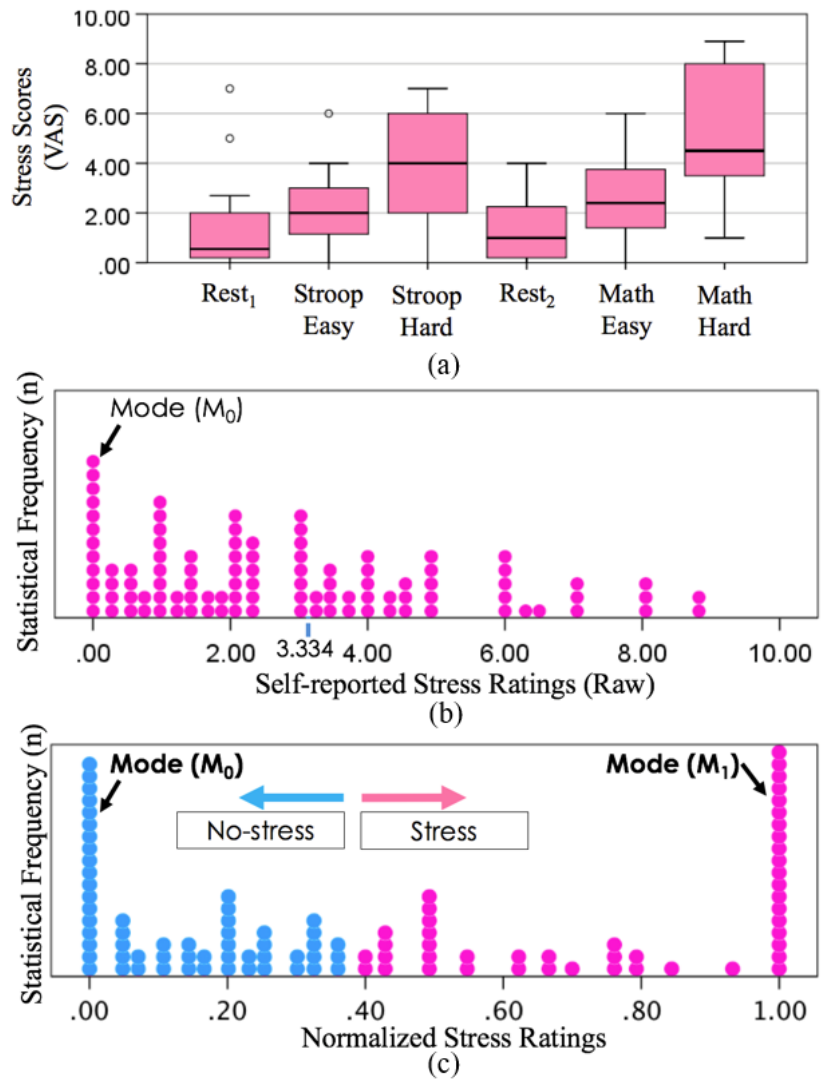

Figure 8. (a) Inter-subject variability shown from the original self-reported stress scores of the 17 participants (box plot, 95% confidence interval) across each section (Rest$_1$, Stroop Easy, Stroop Hard, Rest$_2$, Math Easy, Math Hard). (b) Overall self-reported stress score distributions (from 17 participants over the sessions including the resting periods), (c) normalized stress scores (normalization of scores from each participant) clustered into No-stress and Stress groups along with outputs of K-means.

We tested the correlations between the original self-reported scores, normalized self-reported scores and the high-level hand-crafted PRV and thermal metrics as summarized in Table 1 (using Pearson correlation coefficients). The normalized self-scores maintained a high correlation with the original scores (r=0.752, p<0.001). While some metrics of each physiological sensing channel were significantly correlated between themselves (e.g. PRV F2 - F4: r=0.838, p<0.001; Thermal F1 - F3: r=0.803, p<0.001), the correlation values were lower across sensing channels. In addition, only SDSTV shows approaching significance but low correlation with the self-report scores (r = .196, p=0.059), indicating that each individual engineered metric alone could not lead to high discrimination between perceived levels of stress.

Table 1. Pearson correlation coefficients across self-reports, PRV (PPG derived HRV) and thermal metrics (High-level features). S1=Normalized self-reported scores, S2=original self-reported scores.

| | | | Self reports | | PRV (PPG derived HRV) | | | | | | Nose Temperature | | |
|---|---|---|---|---|---|---|---|---|---|---|---|---|---|
| | | | S1 | S2 | LF (F1) | HF (F2) | LF/HF (F3) | SDPP (F4) | RMSSD (F5) | pPP50 (F6) | TD (F1) | SDSTV (F2) | SDTV (F3) |
| Self Rep. | S1 | Corr. | 1 | .752 | 0.007 | 0.011 | -0.044 | 0.03 | 0.146 | 0.058 | -0.154 | 0.196 | 0.02 |
| | | p-value | | <0.001 | 0.943 | 0.911 | 0.665 | 0.77 | 0.148 | 0.569 | 0.139 | 0.059 | 0.848 |
| | S2 | Corr. | | 1 | -0.079 | -0.044 | -0.082 | -0.002 | 0.083 | 0.097 | -0.153 | 0.197 | 0.032 |
| | | p-value | | | 0.438 | 0.664 | 0.422 | 0.987 | 0.414 | 0.338 | 0.14 | 0.058 | 0.758 |
| PRV (PPG derived HRV) | F1 | Corr. | | | 1 | 0.394 | 0.573 | 0.638 | 0.098 | 0.134 | 0.016 | 0.12 | 0.047 |
| | | p-value | | | | <0.001 | <0.001 | <0.001 | 0.336 | 0.186 | 0.88 | 0.251 | 0.657 |
| | F2 | Corr. | | | | 1 | -0.293 | 0.838 | 0.13 | 0.39 | 0.083 | 0.2 | 0.054 |
| | | p-value | | | | | 0.003 | <0.001 | 0.199 | <0.001 | 0.431 | 0.054 | 0.608 |
| | F3 | Corr. | | | | | 1 | 0.007 | -0.027 | -0.178 | 0.056 | 0.057 | 0.123 |
| | | p-value | | | | | | 0.948 | 0.791 | 0.079 | 0.596 | 0.588 | 0.239 |
| | F4 | Corr. | | | | | | 1 | 0.139 | 0.571 | 0.1 | 0.198 | 0.084 |
| | | p-value | | | | | | | 0.171 | <0.001 | 0.338 | 0.058 | 0.425 |
| | F5 | Corr. | | | | | | | 1 | -0.067 | -0.059 | 0.174 | -0.067 |
| | | p-value | | | | | | | | 0.511 | 0.572 | 0.095 | 0.521 |
| | F6 | Corr. | | | | | | | | 1 | 0.134 | 0.212 | 0.127 |
| | | p-value | | | | | | | | | 0.2 | 0.042 | 0.225 |
| Temperature | F1 | Corr. | | | | | | | | | 1 | 0.213 | 0.803 |
| | | p-value | | | | | | | | | | 0.039 | <0.001 |
| | F2 | Corr. | | | | | | | | | | 1 | 0.487 |
| | | p-value | | | | | | | | | | | <0.001 |
| | F3 | Corr. | | | | | | | | | | | 1 |
| | | p-value | | | | | | | | | | | |

Figure 9 shows values of each pre-crafted metric across the sessions (rest and four stressful events, i.e. Stroop easy/hard and Math easy/hard) and across the labels produced by the labelling technique. As shown in Figure 9a, there was no common pattern found between two easy or hard tasks, although they were designed to induce similar levels of mental stress (e.g. easy: low stress level, hard: high stress level). For example, Thermal F1 appeared to strongly decrease during the Math hard task but not during the Stroop hard task, Thermal F2 increased with the Stroop hard task, but less during the Math hard task. PRV F5 was generally high after both Math easy and hard task sessions than Stroop hard session. This can indicate further that each metric alone from the instant measurement is less likely to contribute to the inference of each session. On the other hand, when we applied our labelling technique, Thermal F1 values grouped into *stress* were generally lower than *no-stress* data as shown in Figure 9b (consistent with findings from literature [38,40,41]).

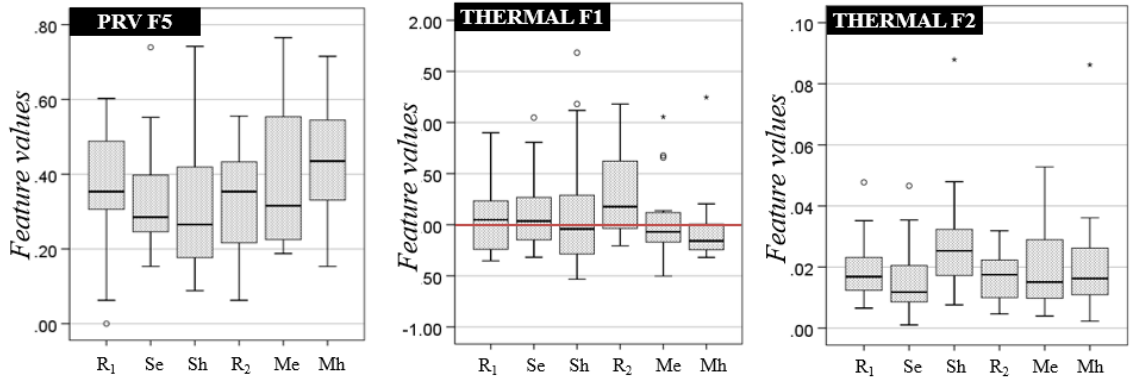

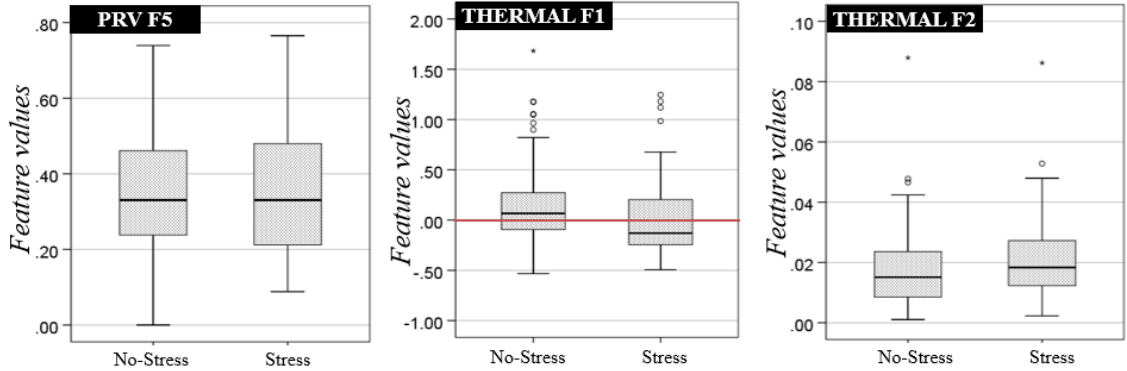

Figure 9. Box plots of 95% confidence intervals in values of each pre-crafted metric across (a) each session ($R_1$: Rest 1, Se: Stroop easy, Sh: Stroop hard, $R_2$: Rest 2, Me: Math easy, Mh: Math hard) and (b) label produced by our labelling technique. The three features (having best correlations with self-reports) are PRV F5:*RMSSD*, root mean square of the successive differences of P-P intervals, Thermal F1:*TD*, temperature difference between from the start and the end (a red line is drawn to show negative or positive thermal direction), F2:*SDTV*, standard deviation of the successive differences of thermal variability sequence.

**Instant Stress Inference Results**
To evaluate the performance of instant stress recognition, we used a 17-fold leave-one-subject (participant)-out (LOSO) cross-validation. LOSO was chosen to test the ability to generalize to unseen participants (*one size fits all*) [36,70]. Figure 10 summarizes the accuracy results of the three classifiers (NN1, NN2, kNN) using LOSO (N=17) for three different cases: a) multimodal-approach by simply combining features from both sensing channels (PRV, Thermal), b) unimodal approach using thermal features, and c) unimodal approach using PRV features. Both neural networks NN1 and NN2 used our proposed low-level features only (i.e. P-P intervals and thermal variability sequences). Overall, the NN2-based multimodal approach produced the highest mean accuracy of 78.33% (SD=15.43) (mean F1 score=77.92%) in discriminating between no-stress and perceived stress (see confusion matrix in Figure 10b for details). The NN1(whose hidden layer is smaller than that for NN2) produced a lower accuracy (M=66.76%, SD=21.75). From all cases of modality, the kNN with the high-level features (i.e. using the hand-engineered 6 PRV and 3 thermal metrics) performed worst. A similar pattern can be seen for the PRV unimodal channel (NN1: M=65.78%, SD=20.55; NN2: M=68.53%, SD=18.89; kNN: M=50.20%, SD=19.63). For the thermal channel, the NN1 appears to perform marginally better (M=58.82%, SD=21.11) than the NN2 (M=56.67%, SD=18.79), but again both NNs perform better than the kNN (M=48.14%, SD=16.52).

However, it should be noted that, for all the models, the confusion matrices for the thermal case (Figure 10b - Thermal) show a clear bias towards the no-stress class. Given this bias and the fact that thermal data from the Rest 1 sessions appeared to be affected by the large variation in temperature between the waiting space and the experiment room (in addition, some participants had just arrived from the outside while others had been already indoor for sometimes), we re-ran the models discarding the data from the Rest 1 sessions. Whilst the overall performance over this modality did not change largely (NN1: M=58.14%, SD=23.33; NN2: M=58.14%, SD=21.59; kNN: M=55.88%, SD=22.38) and the NN1 and NN2 still perform better than the kNN with hand-engineered features, all the confusion matrices (Figure 10c) show more balanced results and a better prediction of the stress class overall.

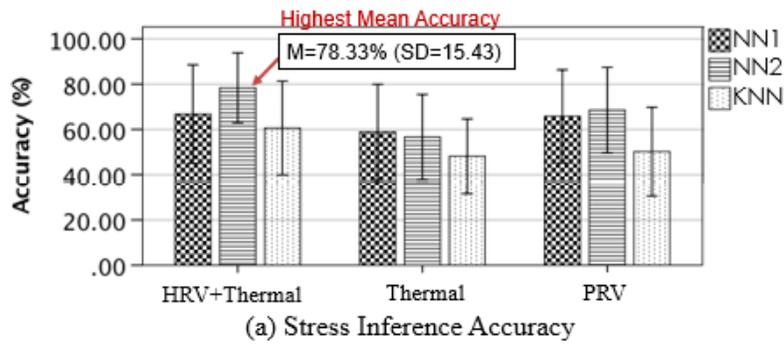
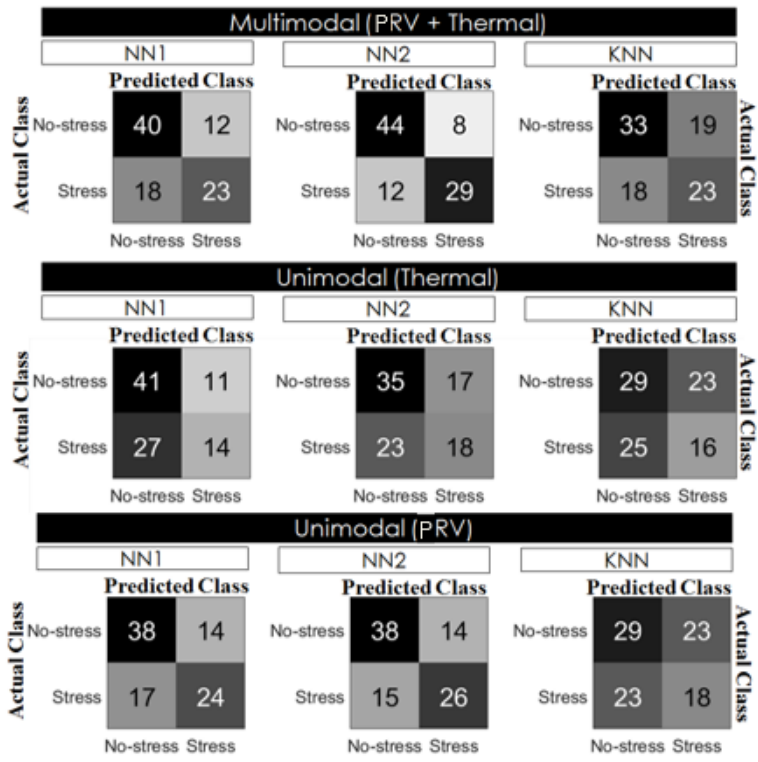
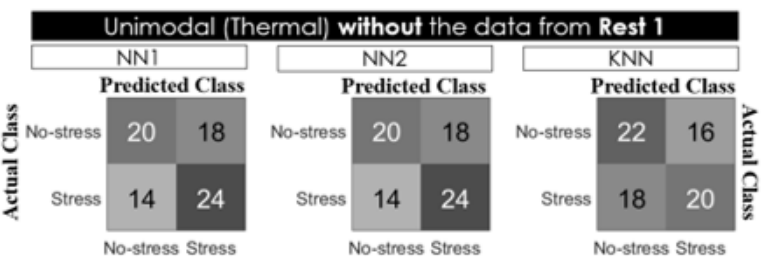

Figure 10. Summary of (a) mean inference accuracy results across 17 folds, (b) accumulated (from 17 LOSO folds) confusion matrices for the three classifiers NN1, NN2 and kNN along with each set of modalities (Multimodal: PRV+Thermal, Unimodal: Thermal, PRV), (c) confusion matrices for the temperature-based unimodal approach built without the Rest 1 data. Each number in the confusion matrices refers to the number of instances.

A repeated measures ANOVA analysis was carried out on results from the 17 folds (including the Rest 1 data) to compare the two NN modeling approaches (that use our proposed low-level features) with the kNN (that uses hand-engineered metrics) to determine whether there was a statistical mean difference in performance. The results show significant differences between the methods for the multi- and the PRV modalities (PRV+Thermal: $F(2,32)= 3.763$, $p=.034$, $\eta_p^2=.190$; PRV: $F(2,32)= 6.001$,

p=.006, $\eta_p^2$=.273). No differences were found for the thermal case (Thermal: F(2,32)= 2.304, p=.116, $\eta_p^2$=.126). Post-hoc paired t-test with Bonferroni correction (see Figure 11) showed that NN2 performed significantly better than kNN for the unimodal PRV case (PRV: p=.023). For the multimodal case, NN2 approached significantly better performance than kNN (PRV+Thermal: p=.064) and NN1 (PRV+Thermal: p=.052). NN1 did not significantly perform better than kNN, however it presented a positive trend in the unimodal PRV case (PRV: p=.091). Even if no significance differences were found over the unimodal thermal case, the graphs in Figure 11 shows how the two NN models performed slightly better than the kNN for all cases including the thermal one. It could be expected that in the case of deployment, a larger sample of data for each class could indeed lead to statistical significance.

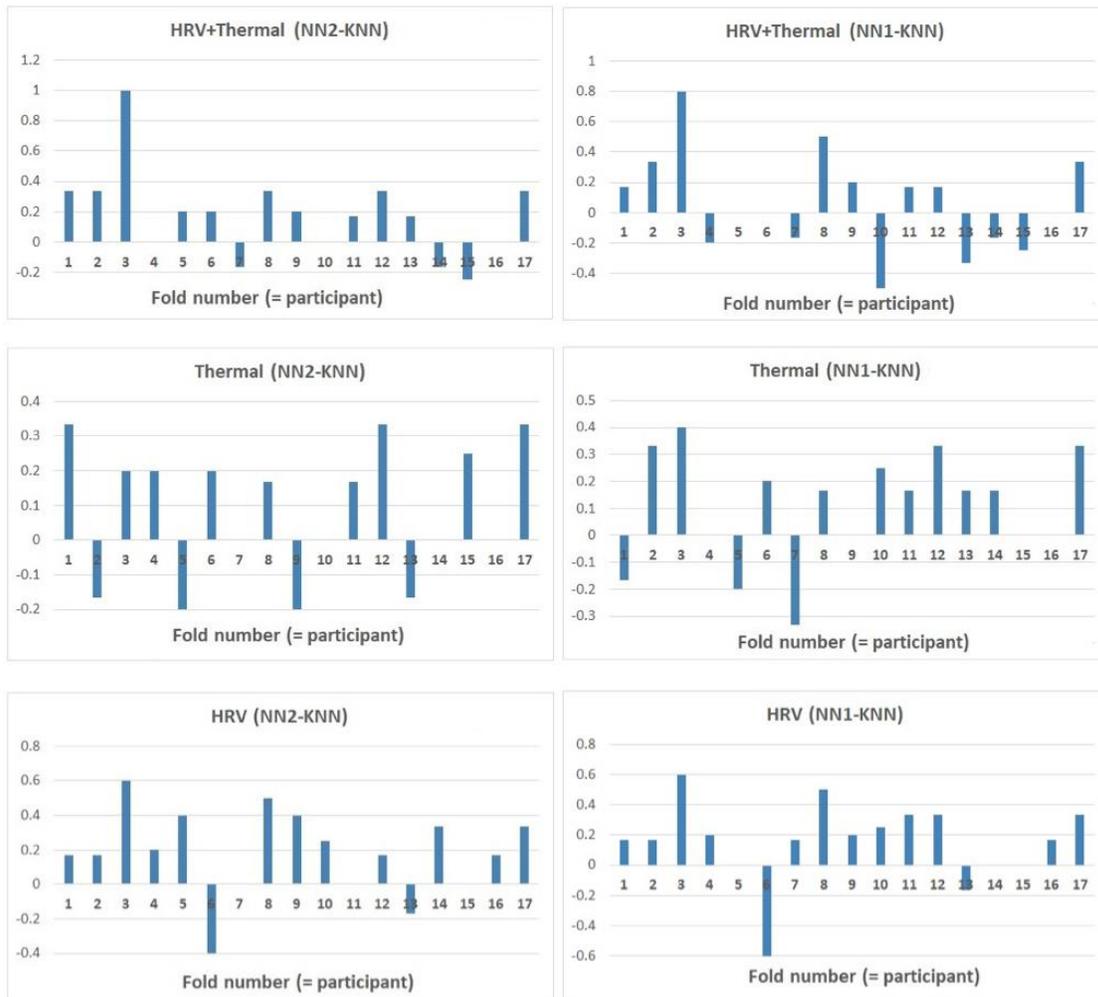

Figure 11. Differences in performance over each fold (i.e. LOSO = for each tested participant data) between the three models over the three modalities. They show how NN2 and to a certain extent NN1 generalize to unseen participants better than kNN.

Lastly, we investigated the effect of the normalization and K-means clustering of self-reported scores in inferring the perceived stress levels. For this part of the study, we removed the Rest1 data. There were two reasons for this. First, we wanted to avoid the noise from the set of data affecting the comparison between the labelling methods. Second, this was also to obtain a more balanced number of instances in each class for testing different labelling methods, less biasing the learning process. The comparison of models over the different labelling techniques did not aim to obtain better performance, but to understand how normalization and different clustering approaches could affect the modeling by acting on class separation and inter-person variability in subjective self-reports. We were also interested in understanding how sensitive the system was in separating stress scores by using the same dataset and merging the intermediate levels with one of the two classes (L1 and L2).

We tested the three models (NN1, NN2, kNN) for the multimodal-approach with the different labelling strategies (L2-L4, introduced in the previous section). Figure 12 summarizes the accuracy results for four different strategies - L1: the main method, L2: K-means with k=3, but combining no-stress and moderate level stress scores as one group, L3: K-means with k=2, dissecting the moderate level scores into no-stress and stress, and L4: original scores divided by a point between no-stress and moderate levels (i.e. 3.334 of 10, see Figure 5c). The results showed that the L1 performed best in separating the bimodal distribution of normalized self-reported scores and helped address the inter-personal variability issue. Indeed, all three models obtained the best accuracy with L1 and the worst performance for L3 and L4 with L4 being marginally better than L3. Finally, it should be noted that in the case of L3 and L4, the best performance was obtained with NN2 rather than NN1. This may indicate that mapping feature values to perceived stress scores may benefit from a larger hidden layer to capture the complexity of the relation.

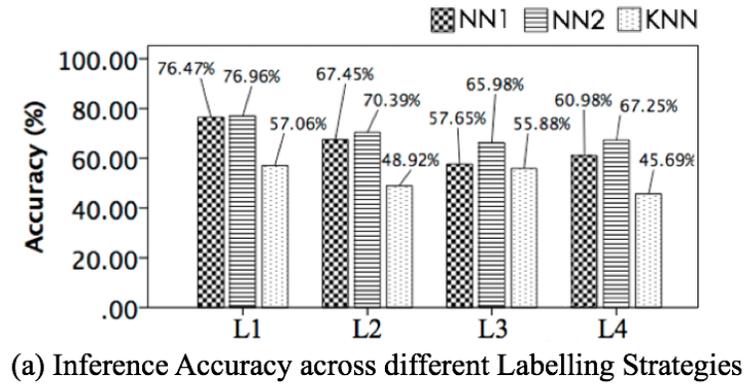

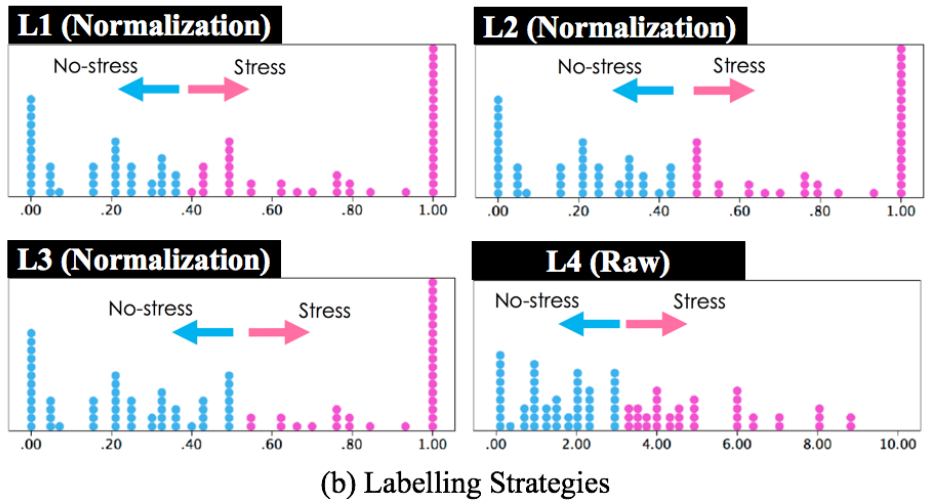

Figure 12. Summary of (a) inference accuracy along with (b) different labelling approaches (L1: K-means with k=3 and combining moderate and high stress scores, L2: K-means with k=3 and combining no-stress and moderate level stress scores, L3: K-means with k=2, L4: original scores divided by the border between no-stress and moderate levels).

## Discussion

This paper contributed to the body of work that aims to make mobile measurements of mental stress more feasible and robust. We focused on two stress-related cardiovascular signals: blood volume pulse and vasoconstriction/dilation-related nose tip temperature. They have been widely investigated in both the mental health and computing literature [22,39,41,47,75], but their applicability together with low cost sensing offered by mobile devices has not been explored. Our work makes four key contributions: (i) a set of methods to improve the quality of the sensed signal, (ii) a demonstration of the limited capability of typically used engineered features in the context of very short-term (*instant*) measurements, (iii) a new set of low level features to capture the dynamical variability of the two signals, (iv) the feasibility of using 20 second measurements to discriminate between no stress and stress responses. Finally, we report on the lesson learned from the analysis of different labelling methods and their effect on the modeling process. Below are detailed discussions of these contributions.

### Towards smartphones as reliable cardiovascular measures

Our first contribution is to develop a new set of preprocessing techniques to enhance the quality of the signal extracted from either the PPG channel (which detects blood pulse variability), or the thermal camera (which detects vasoconstriction/dilation induced nasal temperature variability). This is particularly important in mobile, ubiquitous settings where physiological sensing setups are still of lower quality and have to be less controlled in comparison with the ones generally used in medical environments.

With the data collected from our stress-inducing tasks, we wanted to test the possibility of building algorithms that can reliably and continuously capture i) a person's blood volume pulse pattern from the smartphone camera and ii) nose-tip temperature sequence from the add-on thermal camera. Reliable BVP recording is critical in particular for short term measurements [26,47]. The conducted signal quality test with the pSQI showed that our method produced higher quality BVP signals than the ones obtained with traditional camera-based PPG approaches [6,7,9] (see Figure 6). In addition, we found that a person's respiratory cycles interfered with capturing thermal variations accurately from a person's nasal area (Figure 7). Hence, we built a new technique to minimize such effects and gather a more reliable nose tip thermal signature. This was achieved through the use of an advanced thermal ROI tracking [11] and signal processing techniques to filter out breathing cyclic events (Figure 4) on measured temperatures from the nose area.

However, it should be noted, despite the use of the quantization approach that helps handle environmental temperature changes [11], thermal data during Rest 1 was affected by the difference in temperature between the waiting area and experiment area. This effect was further enhanced when the participants just arrived from outdoors with body temperature being strongly influenced by the cold weather outdoors (winter season). This is important as if the system has to be used, it is crucial for the person to use it in the same environment where stressful events occur. It should also be tested in future studies if a decrease in nose tip temperature may be saturated by very cold environments and therefore be less informative in such situations for automatically detecting mental stress.

### Traditional cardiovascular metrics do not capture stress-related variability from an instant measurement

We found that the capability of the HRV metrics (used as high-level features in the literature [18,20,25]) in instantly quantifying stress was very limited (see Table 1). This is important as despite their general use (e.g. literature in psychology or affective computing), there have still been arguments of such metrics in regards to the possibility of oversimplifying physiological responses [33–35]. It should be noted that although we used PPG-derived metrics rather than the more investigated ECG-derived metrics, strong correlations have been found between the two signal metrics in the case of healthy participants and limited physical movement [29,45]. Stressors in general affect cardiac pulse-related events even if the two types of events (heart rate and blood volume pulse) may be differently affected within non-healthy or elderly population and extreme situations (hot temperature) [75–77]. It should also be noted that although mathematically, a shorter measurement period could lead to a lower resolution of data in the frequency domain resulting in a lower accuracy in computing metrics such as LF/HF [21], recent studies have validated the use of them with very short measurements (10-30s) [47].

Similarly, the metrics applied to short-term nasal thermal data (e.g. TD: Temperature Difference) did also weakly contribute to stress quantification. This may explain inconsistent findings in the literature where such metrics have been used to capture thermal responses to stressful events [41,78]. All in all, the results suggested the need to develop a novel way that describes dynamical information of blood volume pulse and vasoconstriction/dilation-related nasal temperature to help improve the understanding and capturing of their complex phenomenon.

### Overcoming limitations to mobile automatic stress inference

On the basis of the low correlation between perceived mental stress levels and typically engineered metrics for these two signals, we proposed to use thermal variability and P-P interval sequences as a novel set of low-level features to capture stress responses of cardiovascular activities. With this, we investigated how to benefit from automatic feature learning capabilities of machine learning classifiers (i.e. NNs) in instantly inferencing mental stress. The results showed clear improvements in performance. Indeed, our proposed method with the two cardiovascular signals achieved 78.33% correct recognition accuracy with the NN2 whereas only 60.59% from the kNN with the hand-engineered features. Similarly, using the HRV-related features only, there was an improvement by 18.33% with respect to the traditional approach (50.20%). The improvement on the thermal channel was smaller but still evident from the results.

In addition, two further contributions can be highlighted from our approach to the modeling of automatic stress inference: *instant measurements* and *no need for baseline*. First of all, previous work required relatively long-term measurements of between 2 to 5 minutes [25,41,54]. Indeed, our results demonstrated the possibility to use just a 20 seconds measurement to automatically discriminate between stress and non-stress moments. This approach achieved state-of-the-art performance when compared with approaches using much longer measurements (up to around 70 - 80% correct recognition from LOSO cross-validation; e.g. [70]). This is very important given that stillness is critical during PPG measurements and to a certain extent for thermal imaging. In fact, even if automatic ROI tracking methods may help with thermal measurements, people tend to easily move away from the camera or cover their nose with their hands (5 participants did so at least once even for 20 seconds).

Second, our approach (more reliable signal and richer features) led to state-of-the-art results without the use of a baseline. This is critical to everyday life settings as in everyday life such baselines may be difficult to establish. Resting periods just before a stressful event cannot be planned and continuously gathering such measures can be costly, whilst at the same time non-stressful resting periods would also need to be automatically detected. In addition, our data from resting periods shows that such a gold standard resting situation does not exist and environment temperature may change drastically, affecting skin temperature. This could have been due to a lab effect but general everyday life may also have specific effects on the data. Even when using differential features (e.g. temperature differences between two areas of the face - forehead and nose tip), a baseline period was used [42]. The lack of baseline is overcome here by proposing richer features capturing informative physiological variations over time.

### How do we define the ground truth: what is the best approach?

Setting the ground truth is a difficult process when dealing with subjective reports. How to use self-reports to label the data is a critical issue in the field due to their subjectivity. Interpersonal variability has been repeatedly reported as a critical barrier for building stress inference or quantification systems that can generalize across people [24,70]. The inter-subjectivity of self-reports and the need to reduce the number of classes along with types of applications or the size of the dataset require some decisions on how to refine the labels to be taken. In doing so, there is the danger to add noise to the dataset and hence to the modeling process. We explored how different labeling techniques may affect the modeling process.

We proposed to address this problem. The first step was to use a standard normalization technique to take into account personal score ranges over all tasks that aimed to induce a wide range of stress levels (from none to medium to quite high). This transformation led to a bimodal distribution highlighting at least two opposite levels of stress (low and high - see Figure 8c), whilst it still maintained their strong correlation with the original scores ($r=.752$, $p<0.001$). The bimodal distribution is interesting as, given the low number of participants, it suggests the moderate level of stress is not well separated from the other two classes. A binary classification was hence a sensible approach to take in this paper, however with larger datasets a more refined analysis and modeling should be carried out. Second, we used a machine learning clustering technique, K-means, to improve separation of the scores into two classes of stress. The results obtained from the comparison of our approach (L1) with its variation (L2) and the more typically used approaches (L3 and L4) led to an interesting lesson on how to create a more reliable ground truth rather than increase noise in labelling.

Then, how should the data be clustered?: according to the number of stress levels to be recognized, or according to the number of stress levels the data collection experiment was set to induce? The latter approach appeared to be more successful. All labelling methods using K=3 (L1, L2 and to a certain extent L4) performed better than L3 using K=2. This suggests that directly clustering according to the number of classes to be recognized (2 in our case) may spread instances with similar stress level responses (in this case medium responses) across classes introducing noise rather than overcoming the problems of intersubjectivity. However, it should be noted that the normalization step was important. Indeed, the models built on either L1 and L2 using the normalized scored performed better than L4 where the original scores were used instead.

Another important issue to be addressed is: how should the data be grouped when the number of classes to be detected is smaller than the number of levels induced? This decision could be needed either because there were no sufficient instances for a more refined inference or because the application at hand did not require such level of granularity (at the risk of introducing noise due to intersubjective variability). The results showed that L1, collapsing the moderate level with the high one into one class, led to better performance than L2 where medium and no/low stress scores were instead combined. This may suggest that, unless the stress level is very low, stress responses share more similarities than with no-stress responses. A more in-depth analysis of this aspects could be part of a future work and it may require an in-depth analysis of individual responses and validations over other datasets.

Whilst the results provide some interesting insights on how to cluster data from experiments, a question remains on how to deal with data from real-life situations. It is expected that in real-life situations larger datasets may enable finer levels of discrimination personalized to a specific person. In such situations, as the dataset grows, parameters for labelling may need to be adapted to optimize the personalization. However, such rules we used could be helpful to bootstrap models on the basis of experimental datasets or well-structured initial real-life data collections. The bootstrapped models could then be personalized to specific users and recognition levels as data would be continuously collected by the person.

### Limitations and Future Directions

Despite the findings and contributions described above, there is still space for improvement. First, our proposed approach did not perform properly on multiple levels of stress (labelling the data using perceived self-scores). As discussed, this was most probably due to the limited size of the dataset, especially for the medium level of stress (out of three levels). Deploying built software in real life could be a way to build a larger dataset. With a function to collect self-reported person's perceived stress scores (e.g. digitalized VAS sliding bar in an app), this data collection in the wild could produce a sufficient size of cardiovascular signals sets to support more reliable performance in inferencing multiple levels. In addition, it would be interesting to investigate how the transformation of the self-reported scores could be used to support multi-class classification.

Secondly, this work focused on sedentary situations (but without constraining one's mobility) and did not include physical activity (e.g. walking). It is well known that physical activity induces cardiovascular changes, in turn affecting stress inference performance (e.g. [58]). Hence, it would be interesting to test the instant stress inference ability of our system in situations where there is a considerable amount of physical activities, for example, industrial factory work floor.

Finally, investigating the reliability of mobile sensing technologies themselves was outside the scope of this paper (see reviews on this topic - e.g. [79]). We aimed to contribute a better stress inference method that can be used independently regardless of what sensing technology is used. This may be even more crucial when the sensing technology is may not be as accurate and fine grain as more expensive and medically approved technology is.

## Conclusions

With the long term aim of building a stress monitoring system for mobile, everyday use, this paper focuses on the use of smartphone-based imaging capabilities: PPG and thermal imaging. To overcome the difficulties in using smartphone imaging for long period measurements, we propose a novel method that quickly infers a person's perceived level of stress from instant physiological measurements. This is achieved by i) developing a more reliable PPG sensing technique to extract a person's blood volume pulse and its variability; ii) building a thermal imaging-based vasoconstriction monitoring system; iii) investigating the performance of widely used high-level features from PPG and nasal temperature in instant inference tasks; and then iv) proposing novel low-level features to represent heart rate variability and thermal variability; v) building an automatic feature learning-based multimodal perceived stress recognizer; and vi) investigating effects of clustering self-report scores to take into account the subjectivity of self-reports and ensure clear separation between the level of stress to be modelled.

Through the data collection study with 17 participants and a series of stress inducing tasks with different levels, we demonstrated how this system was able to achieve state-of-the-art performance using 20 seconds of data, rather than 2 to 5 minutes typically required by existing methods. This work makes smartphone imaging-based physiological computing capabilities more feasible for real-life applications opening new possibilities for the development of mental-stress support apps and research.

## Acknowledgements
We thank all the participants participated in our experiment. Youngjun Cho was supported by University College London Overseas Research Scholarship (UCL-ORS) awarded to top quality international postgraduate students.

## Conflicts of Interest
none declared

## Appendix

### I. Single hidden-layer Neural Network (NN)

The single hidden-layer NN is one of the most simplistic models amongst pattern recognition frameworks which are built on the multilayer perceptron (inspired by biological systems). It has only one hidden layer and output layer. The output of node $j$ in each layer can be computed with a nonlinear activation function (e.g. sigmoid, ReLU):

$$y_j(\mathbf{x}, \mathbf{w}) = f\left(\sum_{i=1}^{M} x_i w_{ji}^{(L)} + b_j^{(L)}\right). \tag{4}$$

where $x_i$ is the input variable, $i = 1, \ldots, M$ (the total number of inputs), $b_j^{(L)}$ and $w_{ji}^{(L)}$ are the coefficients called *bias* and *weight*, respectively, in the (L)th layer of the network. The weights are updated during training. The weights are tuned with *error backpropagation* to minimize training errors.

## II. k-Nearest Neighbor (kNN)

The kNN is one of the nonparametric pattern recognition algorithms to solve classification and regression problems. With a positive integer K (practically, odd number such as 1), the algorithm seeks for the K points from the training set, which are closest to a new data point x. Following this, it classifies the new point to the class which has the largest number of points within the K points.